\documentclass[aps,prl,twocolumn,superscriptaddress]{revtex4-2}
\usepackage{amsmath}
\usepackage{mathtools}
\usepackage{amssymb}
\usepackage{mathrsfs}
\usepackage{dsfont}
\usepackage[pdftex,colorlinks=true,urlcolor=blue,linkcolor=blue,citecolor=blue,breaklinks=true,bookmarks=false,hypertexnames=false]{hyperref}
\usepackage{graphicx}
\usepackage{float}
\usepackage{xcolor}

\usepackage{multibib}

\newcommand{\push}{\hspace{0.05cm}}
\newcommand{\pull}{\hspace{-0.05cm}}

\begin{document}

\title{Generating Symmetry-Protected Long-Range Entanglement in Many-Body Systems}

\author{Shovan Dutta}
\affiliation{T.C.M. Group, Cavendish Laboratory, University of Cambridge, JJ Thomson Avenue, Cambridge CB3 0HE, United Kingdom\looseness=-1}
\affiliation{Max Planck Institute for the Physics of Complex Systems, 01187 Dresden, Germany\looseness=-1}
\author{Stefan Kuhr}
\affiliation{Department of Physics, University of Strathclyde,  Glasgow G4 0NG, United Kingdom}
\author{Nigel R. Cooper}
\affiliation{T.C.M. Group, Cavendish Laboratory, University of Cambridge, JJ Thomson Avenue, Cambridge CB3 0HE, United Kingdom\looseness=-1}
\affiliation{Department of Physics and Astronomy, University of Florence, Via G. Sansone 1, 50019 Sesto Fiorentino, Italy\looseness=-1}

\date{\today}

\maketitle

{\bf Entanglement between spatially distant qubits is perhaps the most counterintuitive and vital resource for distributed quantum computing \cite{horodecki2009quantum, wehner2018quantum}. However, despite a few special cases \cite{di2008nested, alkurtass2014optimal, pocklington2021stabilizing, dutta2020long, lotkov2021floquet}, there is no known general procedure to maximally entangle two distant parts of an interacting many-body system. Here we present a symmetry-based approach, whereby one applies several timed pulses to drive a system to a particular symmetry sector with maximal bipartite long-range entanglement. As a concrete example, we demonstrate how a simple sequence of on-site pulses on a qubit array can efficiently produce any given number of stable nonlocal Bell pairs, realizable in several present-day atomic and photonic experimental platforms. More generally, our approach paves a route for novel state preparation by harnessing symmetry. For instance, we show how it enables the creation of long-sought-after superconducting $\eta$ pairs \cite{yang1989eta, moudgalya2020eta, kaneko2019photoinduced} in a repulsive Hubbard model.}

Since the early days of quantum mechanics, entanglement has been seen as a fundamental quantum trait, which now lies at the heart of quantum information processing \cite{horodecki2009quantum, wehner2018quantum}. In recent decades, experiments have made great progress in generating entanglement between two isolated qubits via photon exchange \cite{hofmann2012heralded, kurpiers2018deterministic, humphreys2018deterministic} and combining such two-qubit gates to build multi-qubit states \cite{barmettler2008quantum, mooney2021whole, bluvstein2021quantum}. However, it is much harder to entangle a pair of distant qubits in a truly many-body environment where multiple qubits interact with one another \cite{kendon2002typical}. Theoretical protocols have largely focused on creating a single Bell pair between two ends of a spin chain \cite{wang2010robust, sainz2011entanglement, bayat2010entanglement, estarellas2017robust}. Although more pairs can be entangled using specially designed nonuniform coupling \cite{di2008nested, alkurtass2014optimal} or correlated dissipation \cite{pocklington2021stabilizing}, these are seldom realized in experiments \cite{pitsios2017photonic} and limited to free-fermionic chains.

\begin{figure}
	\includegraphics[width=1\columnwidth]{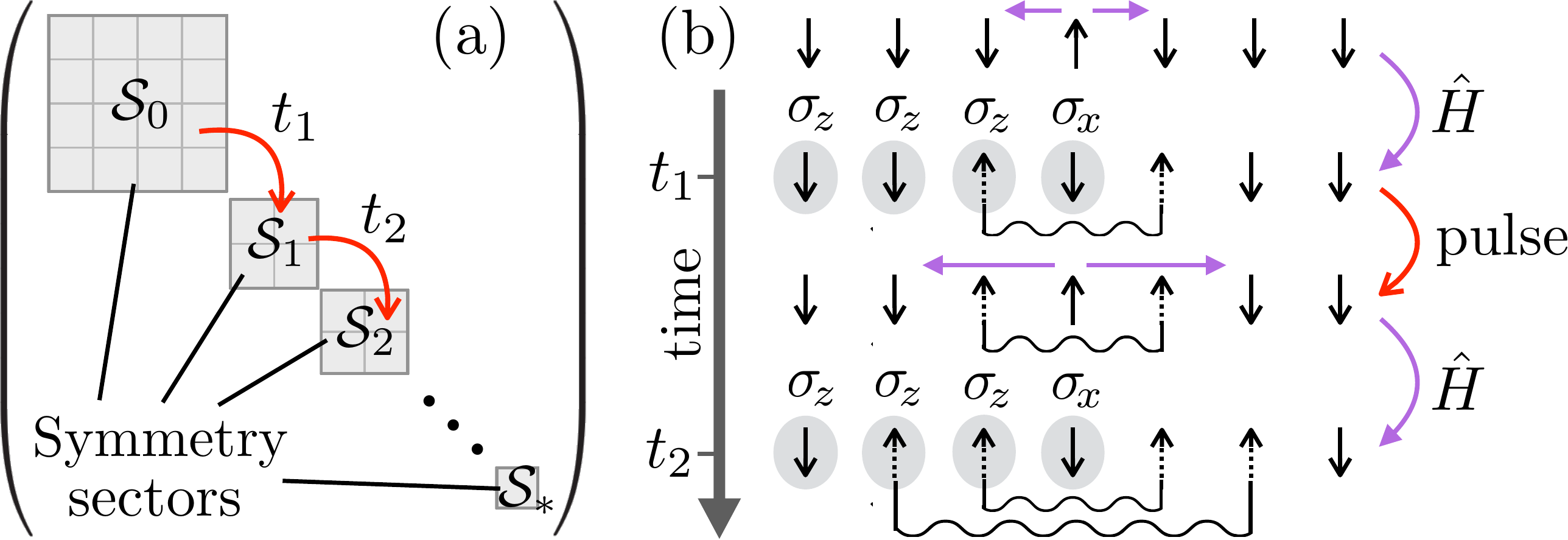}
	\centering
	\caption{{\bf Sketch of the protocol and its application to a qubit array.} (a) Schematic of how one can sequentially transfer a many-body system through symmetry sectors $\mathcal{S}_n$ to prepare a maximally-entangled state in $\mathcal{S}_*$ by a sequence of pulses at optimal times $t_n$. The block-diagonal structure represents the undriven Hamiltonian $\hat{H}$. (b) Illustration for a symmetric spin-1/2 chain: $\pi$ pulses create $\uparrow$ spins at the center site, which then spread under the Hamiltonian, giving rise to multiple Bell pairs between mirror-conjugate sites.
	}
	\label{fig:protocol_sketch}
\end{figure}

Here, we introduce a general approach that leverages the existence of a symmetry to create entanglement over increasingly longer distances. Symmetry plays a fundamental role in modern physics \cite{gross1996role}. As in classical mechanics, a symmetry in a quantum system is intimately linked to a conservation law, which divides the space of all states into decoupled sectors. We envision scenarios where these sectors can be characterized by increasing long-range entanglement, as in the examples below. Our idea is sketched in Fig.~\ref{fig:protocol_sketch}(a): Suppose the system is initially in some low-entangled state in sector $\mathcal{S}_0$, e.g., the ground state or a product state. We apply a series of symmetry-breaking pulses to drive the system toward the maximally-entangled sector $\mathcal{S}_*$, hosting one or few states. The pulses are timed optimally to maximize this one-way transfer. If the transfer fidelity is high, one can produce a macroscopic weight in $\mathcal{S}_*$ with a small number of pulses.

We will show how this technique allows one to generate Bell pairs between any number of mirror-conjugate sites in a uniform spin chain [Fig.~\ref{fig:protocol_sketch}(b)], realizing variants of the so-called ``rainbow'' states \cite{vitagliano2010volume, ramirez2014conformal}. These states possess a high persistency of entanglement \cite{briegel2001persistent} and can be used to efficiently distribute entanglement in a quantum network \cite{di2008nested}. Unlike other protocols for creating rainbow-like states, we do not require the spin-spin interactions to be selectively switched off \cite{barmettler2008quantum}, individually fine tuned \cite{vitagliano2010volume, di2008nested, alkurtass2014optimal, pitsios2017photonic}, or coupled with engineered dissipation for slow relaxation \cite{dutta2020long, pocklington2021stabilizing}. Instead, our scheme uses only local $\pi$ pulses that are standard in experiments. The Bell pairs are created in a time linear in system size, and stable thereafter due to the symmetry conservation.

Crucially, our approach extends beyond spin chains to arbitrary interacting systems with a similar symmetry structure. In particular, we will show how one can drive a Fermi-Hubbard chain \cite{essler2005one} toward an elusive maximally-correlated $\eta$-pairing state \cite{yang1989eta}. Further, since our protocol is symmetry based, the entanglement generated is robust against any perturbations that preserve the symmetry.

{\it Qubit-array protocol.---}We consider the simplest spin-1/2 XX Hamiltonian with $2l+1$ sites for integer $l$,
\begin{equation}
    \hat{H} = -(J/4) \sum\nolimits_{i =-l}^{l-1} \hat{\sigma}^x_i \hat{\sigma}^x_{i+1} + \hat{\sigma}^y_i \hat{\sigma}^y_{i+1} \push,
    \label{eq:hamiltonian}
\end{equation}
where the $\hat{\sigma}$'s are the Pauli spin operators and $J$ is the spin-spin coupling ($\hbar=1$). This model can be reduced to free fermions through a Jordan-Wigner (JW) map, $\hat{f}_i = \hat{\sigma}^-_i \prod_{j<i} \hat{\sigma}^z_j$, where $\hat{\sigma}^{\pm}_i := (\hat{\sigma}^x_i \pm {\rm i} \hat{\sigma}^y_i)/2$ and $\hat{f}_i^{\dagger}$ creates a fermion at site $i$, yielding $\hat{H} = -(J/2) \sum_{i} \hat{f}_i^{\dagger} \hat{f}_{i+1}^{\phantom \dagger} + \;\text{h.c.}$. Thus, spins $\uparrow$ and $\downarrow$ correspond to filled and empty sites, respectively, in the fermion picture, and the occupation of each fermionic single-particle mode is conserved.

In particular, as shown in Ref.~\cite{dutta2020long}, there is a crucial symmetry relating to the entanglement between left and right halves of the chain, given by the symmetry operator
\begin{equation}
    \hat{C} = \hat{\sigma}^z_0/2 + \sum\nolimits_{i=1}^l (\hat{f}_i^{\dagger} \hat{f}_{-i}^{\phantom \dagger}  + \;\text{h.c.}) \push .
    \label{eq:C}
\end{equation}
Here, $\hat{\sigma}^z_0$ measures the fermion-number parity at the center site, and the terms $\hat{f}_i^{\dagger} \hat{f}_{-i}^{\phantom \dagger}$ exchange spins between mirror-conjugate sites $i$ and $-i$ with a phase given by the total parity in between. The nature of this exchange becomes clearer if one rewrites Eq.~\eqref{eq:C} as $\smash{\hat{C}} = \hat{\sigma}^z_0/2 + \sum_{i=1}^l \smash{\hat{a}_{i,+}^{\dagger} \hat{a}_{i,+}^{\phantom \dagger}} - \smash{\hat{a}_{i,-}^{\dagger} \hat{a}_{i,-}^{\phantom \dagger}}$, where $\hat{a}_{i,\pm} := (\smash{\hat{f}_{i} \pm \hat{f}_{-i}})/\sqrt{2}$. Here, $\hat{a}_{i,\pm}^{\dagger}$ creates a Bell pair between sites $i$ and $-i$, $\smash{\hat{a}_{i,\pm}^{\dagger}} |\text{vac}\rangle \propto |\pull\pull\uparrow_i \downarrow_{-i} \rangle \pm |\pull\pull\downarrow_i \uparrow_{-i} \rangle$, where $|\text{vac}\rangle$ is the vacuum state with all spins $\downarrow$. The symmetric and antisymmetric Bell pairs can be thought of as having a ``charge'' of $\pm 1$, such that $\hat{C}$ measures the net ``charge'' of all such pairs. Conversely, this ``charge'' quantum number is 0 for the states $|\pull\pull \downarrow_i \downarrow_{-i} \rangle$ and $|\pull\pull \uparrow_i \uparrow_{-i} \rangle = \hat{a}_{i,+}^{\dagger} \hat{a}_{i,-}^{\dagger} |\text{vac}\rangle$, so it {\it counts} the number of Bell pairs at $\{i,-i\}$ up to a sign. Therefore, the eigenvalues of $\hat{C}$ vary from $\lambda=-l-1/2$ to $\lambda=l+1/2$ in steps of 1, where the $\pm 1/2$ arises from $\hat{\sigma}^z_0/2$. In each of the sectors $\lambda=\pm (l + 1/2)$, there is a single state that is maximally entangled with a Bell pair at all positions, given by $|\Psi_{\pm}\rangle \propto |\pm\rangle_0 \prod_{i=1}^l  |\pull\pull\uparrow_i \downarrow_{-i} \rangle + (-1)^i |\pull\pull\downarrow_i \uparrow_{-i} \rangle$, where $|+\rangle \equiv |\pull\pull\uparrow\rangle$, $|-\rangle \equiv |\pull\pull\downarrow\rangle$, and the alternating phase $(-1)^i$ originates from the JW string. Smaller absolute values of $\lambda$ correspond to states with fewer Bell pairs on average, and all product states have $\langle \hat{C} \rangle = \pm 1/2$.

Our strategy is to start from the vacuum state with $\lambda=-1/2$ and apply a sequence of timed pulses to reach the state $|\Psi_{+}\rangle$. For this, one can increase $\lambda$ by 1 by flipping the center spin from $\downarrow$ to $\uparrow$ [see Eq.~\eqref{eq:C}]. However, one has to ensure it does not affect the exchange terms $\hat{f}_i^{\dagger} \hat{f}_{-i}^{\phantom \dagger}$ which also depend on the center spin. Thus, instead of a $\pi$ pulse at the center site alone, one has to apply a fermionic pulse, e.g., $\hat{f}_0^{\phantom \dagger} + \hat{f}_0^{\dagger} = \hat{\sigma}^x_0 \prod_{i<0} \hat{\sigma}^z_i$, which commutes with $\smash{\hat{f}_i^{\dagger} \hat{f}_{-i}^{\phantom \dagger}}$ $\forall i \neq 0$. This amounts to applying simultaneous local $\pi$ pulses on half the qubits, which can be readily implemented in experiments. Under such a fermionic pulse, $\langle \hat{C} \rangle \to \langle \hat{C} \rangle - \langle \hat{\sigma}^z_0 \rangle$, so if the center spin is $\downarrow$ when the pulse is applied, $\langle \hat{C} \rangle$ increases by 1.

Figure~\ref{fig:protocol_sketch}(b) shows the resulting protocol: Flipping the center spin of the vacuum state gives $\lambda=1/2$ and produces a spin-$\uparrow$ impurity that spreads out in both directions \cite{fukuhara2013quantum}, entangling those sites. During this spreading, $\lambda$ is unaltered as $[\hat{H},\hat{C}]=0$. After a time $t_1 \sim 2/J$, the center site points $\downarrow$ again, when we apply the next pulse, producing another $\uparrow$ spin and increasing $\lambda$ by 1. Repeating this process $l$ times gets one to the state $|\Psi_+\rangle$ with $l$ Bell pairs in a time $t_l \sim 2l/J$.

\begin{figure}
	\includegraphics[width=1\columnwidth]{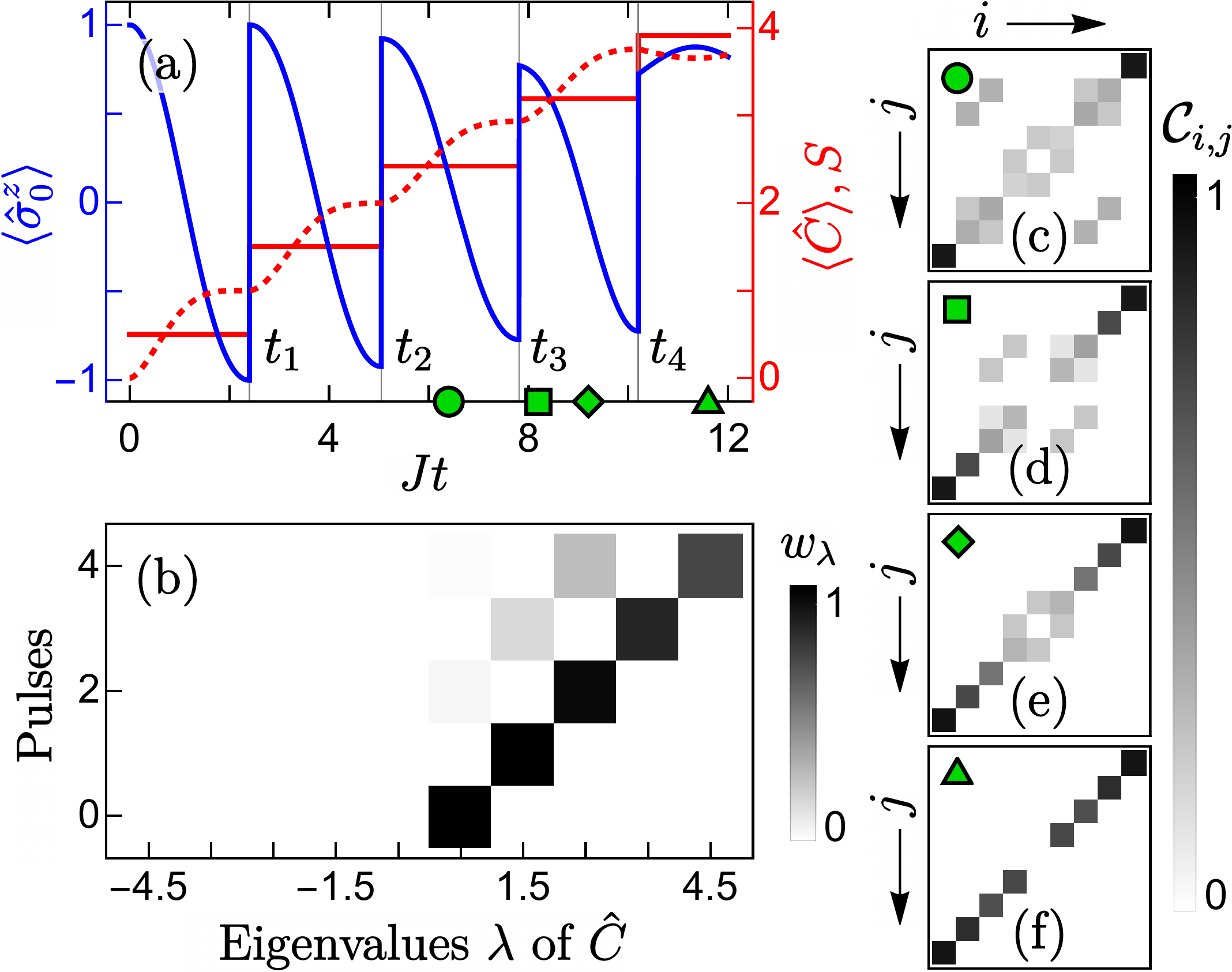}
	\centering
	\caption{{\bf Generation of multiple nonlocal Bell pairs.} (a) Time evolution of the center-spin magnetization $\langle \hat{\sigma}^z_0 \rangle$ of a 9-site spin-1/2 XX chain using exact diagonalization. The center spin is flipped whenever $\langle \hat{\sigma}^z_0 \rangle$ reaches a minimum below 0, which increases the expectation of the symmetry $\smash{\hat{C}}$ (horizontal lines) that relates to the number of Bell pairs between mirror-conjugate sites. Accordingly, the von Neumann entropy $S$ between two sides of the chain grows monotonically (dotted curve). (b) Weights $w_{\lambda}$ in different symmetry sectors after each pulse, showing a high-fidelity transfer toward more strongly-entangled states. (c-f) Pairwise concurrence $\mathcal{C}_{i,j}$ between sites $i$ and $j$ (with $\mathcal{C}_{i,i} \equiv 0$) at times indicated by the green labels in (a), showing how the Bell pairs at $\{i,-i\}$ are successively stacked inward until all positions are filled.
	}
	\label{fig:protocol_numerics}
\end{figure}

The protocol is not perfect as the center does not fully relax to a spin-$\downarrow$ state after the second pulse, as shown in Fig.~\ref{fig:protocol_numerics}(a), where we apply the following pulses whenever $\langle\hat{\sigma}^z_0\rangle$ is minimum (and negative) for the first time. Due to the spin-$\uparrow$ component, there is a small likelihood that $\lambda$ {\it decreases} by 1 during the spin flip. However, as shown in Fig.~\ref{fig:protocol_numerics}(b), the distribution remains strongly peaked at the target sector $\lambda = n+1/2$ after the $n$-th pulse. As a result, both $\langle\hat{C}\rangle$ and the entanglement entropy between left and right halves are near maximal at the end of the sequence [see Fig.~\ref{fig:protocol_numerics}(a)]. Furthermore, the entanglement is distilled in the form of Bell pairs, which is unusual even for highly-entangled states \cite{kendon2002typical}. Such Bell pairs between mirror-conjugate sites are seen in Fig.~\ref{fig:protocol_numerics}(f) where we plot the concurrence between sites $i$ and $j$, which is a robust measure of entanglement between two qubits \cite{wootters1998entanglement} that increases from 0 when the qubits are not entangled to 1 when they are maximally entangled.

Figure~\ref{fig:protocol_numerics}(c) shows the first of these Bell pairs is established between the two end sites when the first $\uparrow$ spin arrives at $t \sim l/J$. Each subsequent pulse adds one more Bell pair toward the center [Figs.~\ref{fig:protocol_numerics}(d-f)]. This stacking is also evident in the experimentally measurable spin-spin correlations (see Extended Data Fig.~\ref{suppfig:spincorr}).

{\it Finite-size scaling.---}As $l$ is increased, the fidelity decreases gradually and saturates for $l \gtrsim 10$, where $\langle \hat{C} \rangle \approx 0.6l + 1$ after the pulse sequence (Extended Data Fig.~\ref{suppfig:scaling}), i.e., the number of generated Bell pairs grows linearly with the number of qubits, as does the preparation time $t\sim 2l/J$. This scaling is on par with more complex protocols using two-qubit gates \cite{barmettler2008quantum} or nonuniform coupling \cite{di2008nested, alkurtass2014optimal}, and much faster than using correlated dissipation \cite{dutta2020long, pocklington2021stabilizing}. The fidelity can be enhanced further by optimal pulse shaping \cite{khaneja2005optimal, li2011optimal}. Moreover, one can project the final state exactly onto $|\Psi_+\rangle$ by measuring the spin imbalance $\hat{\sigma}^z := \sum_i \hat{\sigma}^z_i$, which is locked to $\smash{\hat{C}}$ as $\hat{\sigma}^z = 2(\smash{\hat{C}}-l)$, since $\lambda$ is changed by flipping a spin.

\begin{figure}
	\includegraphics[width=1\columnwidth]{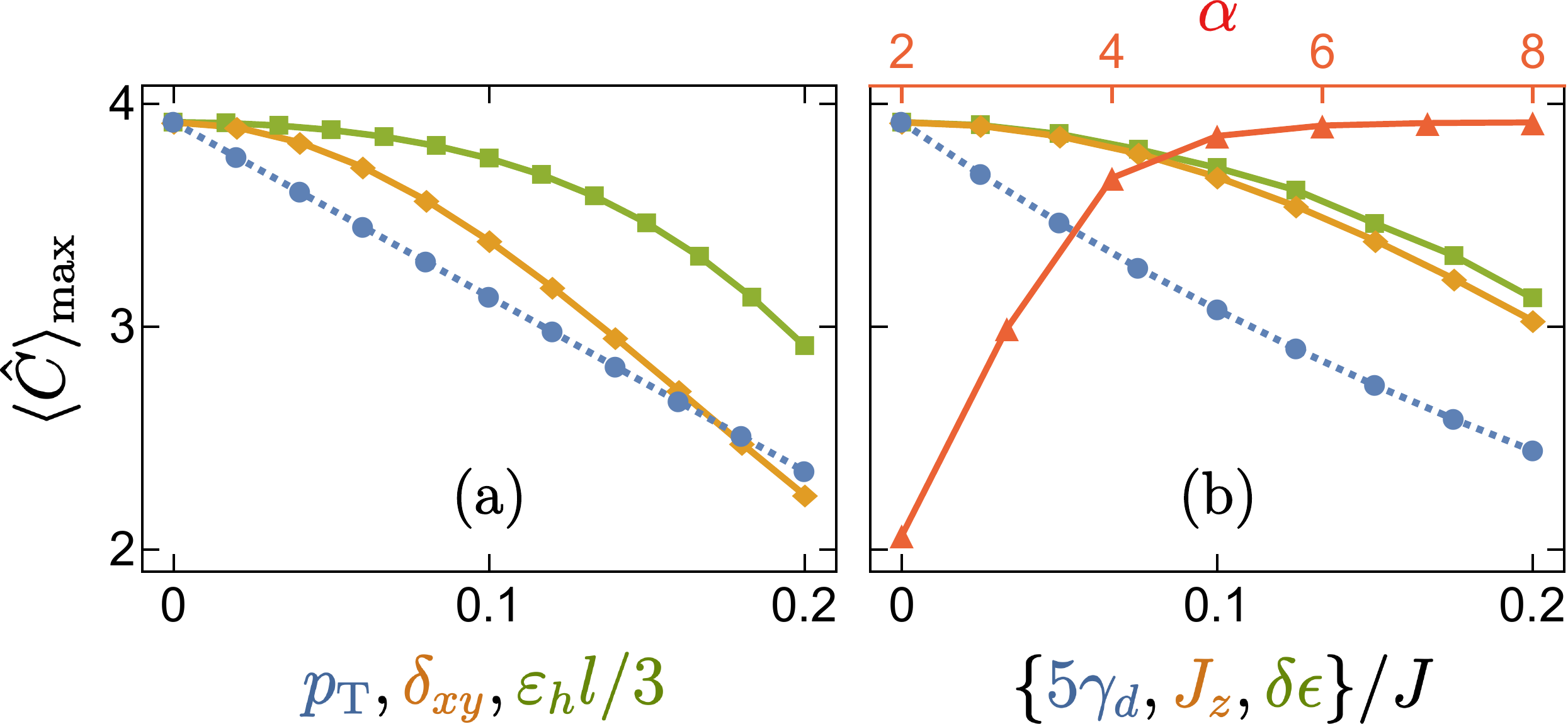}
	\centering
	\caption{{\bf Robustness to generic imperfections}. Fidelity, measured by the maximum value of $\langle\hat{C}\rangle$ attained, for $9$ sites in the presence of (a) symmetry-preserving and (b) symmetry-breaking, unitary (solid) and non-unitary (dotted) perturbations: (a) $p_T$ is a defect probability of starting with an $\uparrow$ spin at any given site; $\delta_{xy}$ is the anisotropy for an XY chain, $\delta_{xy} := (J_x-J_y)/(J_x+J_y)$; $\varepsilon_h$ is a harmonic-trap-induced inhomogeneity, such that $J_{i,i+1} = J/[1-\varepsilon_h^2 (i+1/2)^2]$ \cite{fukuhara2013quantum}; (b) $\gamma_d$ is a uniform dephasing rate, modeled by Lindblad operators $\sqrt{\gamma_d} \hat{\sigma}^z_i/2$ (see Methods); $J_z$ is the $z$-$z$ coupling for an XXZ chain; $\delta\epsilon$ is a disorder strength that gives $\smash{\hat{H}^{\prime}} = \sum_i \epsilon_i \hat{\sigma}^z_i$ where $\epsilon_i$ are randomly distributed in $[-\delta\epsilon/2,\delta\epsilon/2]$; $\alpha$ characterizes long-range interactions $J_{i,j} = J/|i-j|^{\alpha}$. For disorder, dephasing, and thermal defects, $\smash{\langle\hat{C}\rangle}$ is ensemble averaged.
	}
	\label{fig:sensitivity}
\end{figure}

{\it Robustness.---}Figure~\ref{fig:sensitivity} shows the protocol is relatively insensitive to many generic imperfections found in experiments. In particular, the fidelity $\langle\hat{C}\rangle$ is affected only to second order in common Hamiltonian perturbations (see Methods). This includes both symmetry-preserving cases, such as an $x$-$y$ anisotropy, and symmetry-breaking perturbations, such as a $z$-$z$ coupling, random Zeeman splittings, or next-nearest-neighbor coupling.

If a perturbation commutes with both $\hat{C}$ and $\hat{\sigma}^z_0$, the full time evolution of $\langle\hat{C}\rangle$ is unaltered, as in the case of a uniform magnetic field along $z$. For other symmetry-preserving perturbations, including reflection-symmetric potentials \cite{dutta2020long} and dephasing at the center site, the fidelity may be affected but the generated entanglement is stable. If the symmetry itself is broken, $\smash{\langle\hat{C}\rangle}$ attains a maximum value {\it during} the protocol.

{\it Experimental realization.---}The protocol can be implemented on several experimental platforms that have the capability of single-site addressing, particularly quantum-gas microscopes \cite{gross2017expQGM}, superconducting circuits \cite{blais2021circuit}, and arrays of Rydberg atoms \cite{browaeys2020expRydberg}.

Quantum-gas microscope experiments allow us to flip the spin of individual atoms in a chain with high accuracy using a tightly focused laser beam and a microwave field \cite{fukuhara2013quantum}. Using two-component bosons at unit filling in the limit of strong interactions realizes a Heisenberg chain through virtual spin exchange \cite{duan2003controlling}. By separately tuning the intra- and inter-species interactions, one can set the $z$-$z$ coupling to 0, producing the XX model \cite{jepsen2020spin}.

The XX spin-1/2 chain has also been realized using capacitively coupled transmon qubits in a superconducting circuit \cite{ma2019dissipatively}, where one can perform arbitrary single-qubit rotations \cite{blais2021circuit}. The dominant errors come from on-site dephasing and disorder, which are both hundreds of times smaller than $J$ in present-day setups \cite{ma2019dissipatively}.

Resonant dipole-dipole interactions between Rydberg excitations in a chain of atoms also yields an XX model but with long-range coupling, $J_{i,j} = J/|i-j|^{\alpha}$ with $\alpha=3$ \cite{browaeys2020expRydberg}, which breaks the symmetry $\smash{\hat{C}}$. Nonetheless, one can obtain fidelities of up to 75\% for 9 sites [Fig.~\ref{fig:sensitivity}(b)]. 
For trapped-ion chains \cite{monroe2021expIons}, however, $\alpha$ is typically limited to smaller values, making the protocol less viable.

{\it Generalizations: $\eta$ pairing.---}Having demonstrated our approach for a free-fermionic spin chain, we now consider a genuinely interacting system, namely the celebrated Fermi-Hubbard model \cite{essler2005one} with $2l$ sites,
\begin{equation}
    \hat{H} = \pull\sum_{s=\uparrow,\downarrow} \sum_{i =-l}^{l-1}  (-J \hat{c}_{i,s}^{\dagger} \hat{c}_{i+1,s}^{\phantom \dagger} + \text{h.c.})
    + U \pull\sum_{i=-l}^l \hat{n}_{i,\uparrow} \hat{n}_{i,\downarrow} \push.
    \label{eq:Hubbard}
\end{equation}
Here, $\smash{\hat{c}_{i,s}^{\dagger}}$ creates a fermion with spin $s$ at site $i$, $\hat{n}_{i,s} := \smash{\hat{c}_{i,s}^{\dagger}} \hat{c}_{i,s}^{\phantom \dagger}$ gives the site occupation, $J$ is now the nearest-neighbor tunneling, and $U>0$ is an on-site repulsion. It has been known since the 90s that $\hat{H}$ has eigenstates with long-range superconducting order \cite{yang1989eta}, thanks to an SU(2) symmetry with generators $\hat{\eta}^{-} \pull= \sum_i (-1)^i \hat{c}_{i,\uparrow} \hat{c}_{i,\downarrow}$, $\hat{\eta}^+ \pull = \hat{\eta}^{-\dagger}$, and $\hat{\eta}^z \pull= \sum_i (\hat{n}_{i,\uparrow} + \hat{n}_{i,\downarrow}-1)/2$. Here, $\hat{\eta}^+$ acting on the vacuum $|0\rangle$ creates a bound pair (doublon) with quasimomentum $\pi$ (an $\eta$ pair), which leads to staggered superconducting pair correlations $P_{i,j}=\langle \smash{\hat{c}_{i,\downarrow}^{\dagger} \hat{c}_{i,\uparrow}^{\dagger}} \hat{c}_{j,\uparrow}^{\phantom \dagger} \hat{c}_{j,\downarrow}^{\phantom \dagger} \rangle$. In particular, at half filling ($\hat{\eta}_z = 0$), the correlations are maximal for an $\eta$ condensate $|Y\rangle \propto \smash{(\hat{\eta}^+)^l} |0\rangle$, which gives $P_{i,j} = (-1)^{i+j} l/(4l-2) \; \forall i\neq j$ \cite{yang1989eta}. However, producing an $\eta$ pair costs energy $U$, so $|Y\rangle$ is a highly-excited state that is difficult to engineer in theory \cite{kantian2010eta, kaneko2019photoinduced, kraus2008preparation, zhang2020dynamical} and has not been realized experimentally.

To prepare $|Y\rangle$ using pulses, we note that the number of $\eta$ pairs, $N_{\eta}$, is measured by the symmetry $\hat{\eta}^2 = \hat{\eta}^+\hat{\eta}^- - \hat{\eta}^z +(\hat{\eta}^z)^2$ with eigenvalues $N_{\eta} (N_{\eta} + 1)$, where $N_{\eta} = 0,1,\dots,l$ at half filling. The maximally-correlated $|Y\rangle$ is the only state having $N_{\eta} = l$. Further, as shown in Ref.~\cite{kaneko2019photoinduced}, one can change $N_{\eta}$ by $\pm 1$ by applying the current operator $\smash{\hat{\mathcal{J}}} = {\rm i} \sum_{i,s} \smash{\hat{c}_{i+1,s}^{\dagger}} \hat{c}_{i,s}^{\phantom \dagger} + \text{h.c.}$: This amounts to changing $J \to {\rm i} J^{\prime}(t)$ in Eq.~\eqref{eq:Hubbard} for a short pulse, which would be challenging to implement, but possible via laser-assisted tunneling \cite{goldman2014light} or lattice shaking \cite{eckardt2017colloquium}.

\begin{figure}
	\includegraphics[width=1\columnwidth]{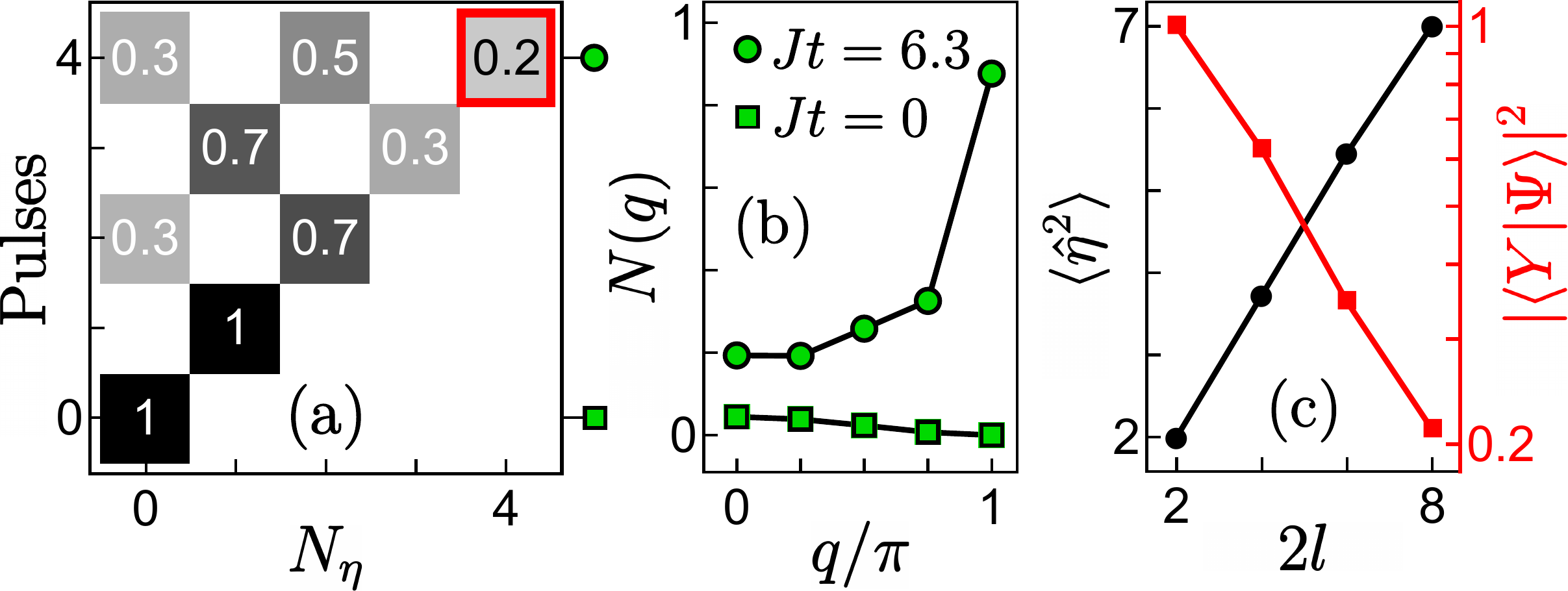}
	\centering
	\caption{{\bf Generation of superconducting $\eta$ pairs.} Evolution of a repulsive Fermi-Hubbard chain ($U/J=10$), starting from an antiferromagnetic ground state at half filling, using exact diagonalization. (a) Distribution among symmetry sectors of $\hat{\eta}^2$ with different numbers of $\eta$ pairs, $N_{\eta}$, after successive pulses for $2l=8$ sites. The red square highlights a notable overlap with the maximally-correlated $\eta$ condensate $|Y\rangle$. (b) Initial and final pair-momentum distributions, showing the emergence of a peak at the band edge, characteristic of $\eta$ pairing. (c) The final value of $\langle\hat{\eta}^2\rangle$ grows linearly with $l$, but the many-body overlap with $|Y\rangle$ falls exponentially \cite{kantian2010eta}.
	}
	\label{fig:eta-pairing}
\end{figure}

Thus, we arrive at a conceptually simple protocol: Starting from an antiferromagnetic ground state with $N_{\eta} = 0$ for $U \gg J$, as realized in Ref.~\cite{boll2016spin}, we repeatedly apply $\smash{\hat{\mathcal{J}}}$ when it would increase $\langle\hat{\eta}^2\rangle$ the most. To this end, we monitor $\mathcal{T}:= \langle \smash{\hat{\mathcal{J}} \hat{\eta}^2 \hat{\mathcal{J}}} \rangle / \langle \smash{\hat{\mathcal{J}}^2} \rangle$ and apply the next pulse whenever $\mathcal{T}$ is maximum over a virtual-tunneling time $\Delta t = U/(2 J^2)$, the slowest time scale in the problem (see Extended Data Fig.~\ref{suppfig:etapairing}). Figure~\ref{fig:eta-pairing}(a) shows the state after $l$ pulses has a significant overlap with $|Y\rangle$, which is reached in a time $t \sim 6/J$ for 8 sites, orders of magnitude faster than adiabatic \cite{kantian2010eta} or dissipative \cite{zhang2020dynamical} approaches. The $\eta$ pairing is manifest in the pair-momentum distribution $N(q) = (2l)^{-1}\pull\sum_{i,j} e^{{\rm i}q(i-j)} P_{i,j}$ as a sharp peak at $q=\pi$ [Fig.~\ref{fig:eta-pairing}(b)], which can be detected experimentally \cite{regal2004observation}. We find the preparation time and the final value of $\langle\hat{\eta}^2\rangle$ both grow linearly with $l$  [Fig.~\ref{fig:eta-pairing}(c)], so the average number of $\eta$ pairs grows as $\sqrt{l}$.

The Hubbard model also has a spin-SU(2) symmetry whose generators $\hat{S}^{\pm}$ and $\hat{S}^z$ are related to the $\eta$ generators by a particle-hole transformation $c_{i,\downarrow} \to (-1)^{j+1} \hat{c}_{i,\downarrow}^{\dagger}$ \cite{essler2005one}. Hence, there is a dual protocol for maximizing the total spin $\langle\hat{S}^2\rangle$ for long-range spin-spin correlations, with the pulse operator $\smash{\hat{\mathcal{J}}_s} = {\rm i} \sum_{i} \smash{\hat{c}_{i+1,\uparrow}^{\dagger}} \hat{c}_{i,\uparrow}^{\phantom \dagger} - \smash{\hat{c}_{i+1,\downarrow}^{\dagger}} \hat{c}_{i,\downarrow}^{\phantom \dagger} + \text{h.c.}$, i.e., a spin-dependent complex tunneling as in Ref.~\cite{aidelsburger2013realization}.

{\it Conclusions.---}We have introduced a general technique for harnessing symmetries to produce on-demand long-range entanglement in many-body quantum systems. The approach relies solely on the symmetry structure and can be readily extended to dissipative systems \cite{dutta2020long, nakagawa2021eta, tindall2019heating} and higher dimensions \cite{yang1989eta, moudgalya2020eta}. For simplicity, we have assumed instantaneous pulses; one can obtain even higher fidelities by allowing more general waveforms using optimal control strategies \cite{khaneja2005optimal, li2011optimal}. Together with advances in engineering many-body Hamiltonians~\cite{weitenberg2021tailoring, gross2017expQGM, blais2021circuit, browaeys2020expRydberg, monroe2021expIons} and dissipation \cite{muller2012engineered}, our technique paves an exciting route to synthesizing strongly-entangled quantum states with key applications to quantum information processing.

We thank Leonardo Banchi and Berislav Bu{\v{c}}a for discussions. This work was supported by EPSRC Grant No. EP/P009565/1 and by a Simons Investigator Award.

\begingroup

\endgroup

\begingroup
\fontsize{9}{11}\selectfont

\bigskip
\noindent {\bf\large Methods}

\medskip

{\bf Equation of motion for the qubit array.} Our protocol for generating Bell pairs leads to a closed evolution of the two-point correlations $\langle \hat{f}_i^{\dagger} \hat{f}_j \rangle$ and $\langle \hat{f}_i \hat{f}_j \rangle$, where $\hat{f}_i$ are the JW fermions. In between pulses, they evolve under the free-fermion Hamiltonian, which gives
\begin{subequations}
\begin{flalign}
    \partial_t \langle \hat{f}_i^{\dagger} \hat{f}_j \rangle &= 
    ({\rm i}J/2) \push \big\langle
      \hat{f}_{i}^{\dagger} \hat{f}_{j+1} 
    + \hat{f}_{i}^{\dagger} \hat{f}_{j-1} 
    - \hat{f}_{i+1}^{\dagger} \hat{f}_{j} 
    - \hat{f}_{i-1}^{\dagger} \hat{f}_{j}
    \big\rangle \push, \hspace{-1cm} &&
    \label{suppeq:fdagfeom1} \\[0.1cm]
    \partial_t \langle \hat{f}_i \hat{f}_j \rangle &= 
    ({\rm i}J/2) \push \big\langle
      \hat{f}_{i} \hat{f}_{j+1} 
    + \hat{f}_{i} \hat{f}_{j-1} 
    + \hat{f}_{i+1} \hat{f}_{j} 
    + \hat{f}_{i-1} \hat{f}_{j} 
    \big\rangle \hspace{-1cm} &&
    \label{suppeq:ffeom1}
\end{flalign}
\end{subequations}
$\forall i,j$, with $\hat{f}_{l+1} := \hat{f}_{-l-1} := 0$. A fermionic pulse $\hat{f}_0 + \hat{f}_0^{\dagger}$ changes the expectation $\langle\hat{O}\rangle$ to $\langle (\hat{f}_0^{\dagger} + \hat{f}_0) \hat{O} (\hat{f}_0 + \hat{f}_0^{\dagger}) \rangle$, which couples $\langle \hat{f}_i^{\dagger} \hat{f}_j \rangle$ and $\langle \hat{f}_i \hat{f}_j \rangle$ through the substitutions
\begin{subequations}
\begin{align}
    \langle \hat{f}_0^{\dagger} \hat{f}_0 \rangle \longrightarrow & \;
    1 - \langle \hat{f}_0^{\dagger} \hat{f}_0 \rangle \push, 
    \label{suppeq:sub_center}\\[0.1cm]
    \langle \hat{f}_i^{\dagger} \hat{f}_0 \rangle \longleftrightarrow &\;
    \langle \hat{f}_i \hat{f}_0 \rangle^* \;\; \forall i \neq 0 \push, 
    \label{suppeq:sub_fidagf0}\\[0.1cm]
    \langle \hat{f}_0^{\dagger} \hat{f}_i \rangle \longleftrightarrow &
    - \langle \hat{f}_0 \hat{f}_i \rangle \;\; \forall i \neq 0 \;.
    \label{suppeq:sub_f0dagfi}
\end{align}
\end{subequations}
Such a pulse is applied when $\langle \hat{\sigma}^z_0 \rangle$ reaches a minimum below 0. This can be monitored using $\hat{\sigma}^z_0 = 2 \hat{f}_0^{\dagger} \hat{f}_0 - 1$, which yields
\begin{subequations}
\begin{align}
    \partial_t \langle \hat{\sigma}^z_0 \rangle &= 
    2 J\; \text{Im} \push \big\langle 
    \hat{f}_{-1}^{\dagger} \hat{f}_0 - 
    \hat{f}_{0}^{\dagger} \hat{f}_1 
    \big\rangle \push,
    \label{sigmaz0prime}\\[0.1cm]
    \nonumber \partial_t^2 \langle \hat{\sigma}^z_0 \rangle &= 
    J^2\; \text{Re} \push \big\langle
    \hat{f}_{-1}^{\dagger} \hat{f}_{-1} +
    \hat{f}_{1}^{\dagger} \hat{f}_{1} -
    2 \hat{f}_{0}^{\dagger} \hat{f}_{0} \\
    &\hspace{1.35cm} + 2 \hat{f}_{-1}^{\dagger} \hat{f}_{1} -
    \hat{f}_{-2}^{\dagger} \hat{f}_{0} -
    \hat{f}_{0}^{\dagger} \hat{f}_{2}
    \big\rangle \push,
    \label{sigmaz0prime2}
\end{align}
\end{subequations}
where Re and Im denote real and imaginary parts. Therefore, starting with a single $\uparrow$ spin at $i=0$, we evolve Eqs.~\eqref{suppeq:fdagfeom1} and \eqref{suppeq:ffeom1}, making the changes in Eqs.~\eqref{suppeq:sub_center}--\eqref{suppeq:sub_f0dagfi} whenever $\partial_t \langle \hat{\sigma}^z_0 \rangle = 0$, $\partial_t^2 \langle \hat{\sigma}^z_0 \rangle > 0$, and $\langle \hat{\sigma}^z_0 \rangle < 0$, for up to $l$ times, as shown in Fig.~\ref{fig:protocol_numerics}(a) and Extended Data Fig.~\ref{suppfig:scaling}(a).

\medskip

{\bf Spin-spin correlations.} The Bell pairs can be detected in experiments by measuring the spin-spin correlations $\langle \hat{\sigma}^{\nu}_i \hat{\sigma}^{\nu}_j \rangle$ for $\nu=x,y,z$, plotted in Extended Data Fig.~\ref{suppfig:spincorr}. Since the dynamics are generated by a quadratic Hamiltonian, the many-body state is Gaussian \cite{shi2018variational} and we can find $\langle \hat{\sigma}^{\nu}_i \hat{\sigma}^{\nu}_j \rangle$ from $\langle \hat{f}_i^{\dagger} \hat{f}_j \rangle$ and $\smash{\langle \hat{f}_i \hat{f}_j \rangle}$ using Wick's theorem. To this end, we first define $\hat{A}_i := \hat{f}_i^{\dagger} + \hat{f}_i$ and $\hat{B}_i := \hat{f}_i^{\dagger} - \hat{f}_i$. Substituting $\hat{\sigma}^z_i = \hat{B}_i \hat{A}_i$ in the JW transformation gives, for $i<j$,
\begin{subequations}
\begin{align}
    \langle \hat{\sigma}^z_i \hat{\sigma}^z_j \rangle &= 
    \langle \hat{B}_i \hat{A}_i \hat{B}_j \hat{A}_j \rangle \push,
    \label{suppeq:zz}\\
    \langle \hat{\sigma}^x_i \hat{\sigma}^x_j \rangle &= 
    \left\langle \prod_{k=i}^{j-1} \hat{B}_k \hat{A}_{k+1} \right\rangle \push,
    \label{suppeq:xx}\\
    \langle \hat{\sigma}^y_i \hat{\sigma}^y_j \rangle &= 
    (-1)^{j-i} \left\langle \prod_{k=i}^{j-1} \hat{A}_k \hat{B}_{k+1} \right\rangle \push.
    \label{suppeq:yy}
\end{align}
\end{subequations}
Applying Wick's theorem to Eq.~\eqref{suppeq:zz}, we find
\begin{equation}
    \hspace{-0.17cm}\langle \hat{\sigma}^z_i \hat{\sigma}^z_j \rangle =
    \langle \hat{B}_i \hat{A}_i \rangle \langle \hat{B}_j \hat{A}_j \rangle -
    \langle \hat{B}_i \hat{B}_j \rangle \langle \hat{A}_i \hat{A}_j \rangle +
    \langle \hat{B}_i \hat{A}_j \rangle \langle \hat{A}_i \hat{B}_j \rangle \push.
    \label{suppeq:zzresult}
\end{equation}
Similarly, the string correlation in $\langle \hat{\sigma}^x_i \hat{\sigma}^x_j \rangle$ [Eq.~\eqref{suppeq:xx}] can be reduced to the Pfaffian of an antisymmetric matrix $\mathbf{\Gamma}^x$ \cite{kormos2017inhomogeneous} of size $2(j-i)$, with the elements
\begin{equation}
    \Gamma^x_{m,n} = 
    \begin{pmatrix}
      \langle \hat{B}_m \hat{B}_n \rangle + \delta_{m,n} & 
      \langle \hat{B}_m \hat{A}_{n+1} \rangle \\[0.1cm]
      \langle \hat{A}_{m+1} \hat{B}_{n} \rangle &
      \langle \hat{A}_{m+1} \hat{A}_{n+1} \rangle - \delta_{m,n}
    \end{pmatrix}
    \smallskip
    \label{suppeq:Gamma}
\end{equation}
where $m,n = i,i+1,\dots,j-1$. One can find $\Gamma^x_{m,n}$ using
\begin{subequations}
\begin{align}
    \langle \hat{A}_m \hat{A}_n \rangle 
    &= \delta_{m,n} + 2 {\rm i} \;\text{Im} \push \big\langle \hat{f}_m \hat{f}_n + \hat{f}_m^{\dagger} \hat{f}_n \big\rangle \push,
    \label{suppeq:AA}\\[0.1cm]
    \langle \hat{B}_m \hat{B}_n \rangle 
    &= -\delta_{m,n} + 2 {\rm i} \;\text{Im} \push \big\langle \hat{f}_m \hat{f}_n - \hat{f}_m^{\dagger} \hat{f}_n \big\rangle \push,
    \label{suppeq:BB}\\[0.1cm]
    \langle \hat{A}_m \hat{B}_n \rangle 
    &= \delta_{m,n} - 2 {\rm i} \;\text{Re} \push \big\langle \hat{f}_m \hat{f}_n + \hat{f}_m^{\dagger} \hat{f}_n \big\rangle \push,
    \label{suppeq:AB}\\[0.1cm]
    \langle \hat{B}_m \hat{A}_n \rangle 
    &= - \langle \hat{A}_n \hat{B}_m \rangle \push.
    \label{suppeq:BA}
\end{align}
\end{subequations}
From Eq.~\eqref{suppeq:yy}, $\smash{\langle \hat{\sigma}^y_i \hat{\sigma}^y_j \rangle}$ can be obtained by interchanging $\hat{A}$ and $\hat{B}$ in $\mathbf{\Gamma}^x$ and multiplying the resulting Pfaffian by $(-1)^{j-i}$.

\medskip

{\bf Robustness for Hamiltonian perturbations.} In the presence of a perturbation $\epsilon \hat{H}^{\prime}$, the dynamics are generated by $\hat{U}(t) = e^{-{\rm i}(\hat{H} + \epsilon \hat{H}^{\prime}) t}$. Using the Zassenhaus formula \cite{magnus1954exponential}, we write
\begin{align}
    \hat{U}^{\dagger}(t) = 
    e^{{\rm i} \hat{H} t} \big[1 - \epsilon \hat{P}(t) + \mathcal{O}(\epsilon^2) \big] \push,
    \label{suppeq:Udag}
\end{align}
where $\hat{P}(t)$ is an anti-hermitian operator given by
\begin{align}
    \hat{P}(t) =&\; 
    \sum_{n=1}^{\infty} \frac{(-{\rm i} t)^n}{n!} \big[ (\hat{H})^{n-1}, \hat{H}^{\prime} \big] \push,
    \label{suppeq:perturbation}\\[0.1cm]
    \text{with}\;\; \big[ (\hat{H})^{n}, \hat{H}^{\prime} \big] :=&\;
    [\underbrace{
    \hat{H},\dots [\hat{H},[\hat{H}}_{n\;\text{times}}
    ,\hat{H}^{\prime}]]\dots] \push,
    \label{suppeq:repeatedcom}
\end{align}
and $\smash{\big[ (\hat{H})^0, \hat{H}^{\prime} \big] := \hat{H}^{\prime}}$. Hence, the change in the expectation value of an operator $\hat{O}$ at time $t$ is 
\begin{align}
    \nonumber
    \delta \langle \hat{O} \rangle (t) &= 
    \langle \Psi(0) | \hat{U}^{\dagger}(t) \push \hat{O} \push \hat{U}(t) -
     e^{{\rm i} \hat{H} t} \push \hat{O} \push e^{-{\rm i} \hat{H} t} | \Psi(0) \rangle
    \\[0.1cm]
    &= \epsilon \langle \Psi(t) | [\hat{O}, \hat{P}(t)] | \Psi(t) \rangle + \mathcal{O}(\epsilon^2) \push,
\end{align}
where $|\Psi(t)\rangle$ is the state of the unperturbed system. So, the first-order correction vanishes provided $\langle [\hat{O}, \hat{P}(t) ] \rangle = 0$. From Eq.~\eqref{suppeq:perturbation}, this is true at all times if $\langle [\hat{O}, [(\hat{H})^n,\hat{H}^{\prime}]] \rangle = 0 \; \forall n$. This condition is, in fact, satisfied for the operators $\hat{C}$ and $\hat{\sigma}^z_0$ for a number of common perturbations $\hat{H}^{\prime}$, leading to the quadratic variations in Fig.~\ref{fig:sensitivity}.

\medskip

{\bf Equation of motion with perturbations.} The dynamics remain free fermionic for a number of variations, including nonuniform coupling $J_i$, Zeeman splittings $\epsilon_i$, and dephasing rates $\gamma_i > 0$. The first two are described by the Hamiltonian
\begin{equation}
    \hat{H} = -\frac{1}{2} \sum_{i=-l}^{l-1} \big(J_i \hat{f}_i^{\dagger} \hat{f}_{i+1} + \text{h.c.}\big) +
    \sum_{i=-l}^l \epsilon_i \hat{f}_i^{\dagger} \hat{f}_i \push,
    \label{suppeq:nonuniformH}
\end{equation}
whereas the dephasing can be modeled by Lindblad operators $\hat{L}_i = \sqrt{\gamma_i} \push \hat{\sigma}^z_i/2$ under a standard Born-Markov approximation \cite{carmichael1999statistical}, resulting in the evolution
\begin{equation}
    \partial_t \langle \hat{O} \rangle = {\rm i} \langle [\hat{H}, \hat{O}] \rangle +
    \sum\nolimits_i \big\langle \big[\hat{L}_i, [\hat{O},\hat{L}_i] \big] \big\rangle
    \label{suppeq:lindbladeom}
\end{equation}
for any hermitian operator $\hat{O}$. The dephasing causes local coherences to decay at a rate $\gamma_i$, i.e., $\partial_t \langle \hat{\sigma}^x_i \rangle_L = -\gamma_i \langle \hat{\sigma}^x_i \rangle$, where the subscript $L$ denotes the dephasing component. The correlations $\langle \hat{f}_i^{\dagger} \hat{f}_j \rangle$ and $\langle \hat{f}_i \hat{f}_j \rangle$ again form a closed set of equations, which can be found from Eq.~\eqref{suppeq:lindbladeom} by straightforward algebra.

Similarly, one can include an XY anisotropy, which adds pairing terms $\hat{f}_i \hat{f}_j$ to the Hamiltonian. For other perturbations such as a $z$-$z$ coupling, long-range interactions, or incoherent spin flips, the system is no longer free fermionic and we perform a full many-body simulation.

\clearpage

\onecolumngrid

\setcounter{figure}{0}
\renewcommand{\figurename}{Extended Data Fig.}

\begin{figure}
	\includegraphics[width=0.95\columnwidth]{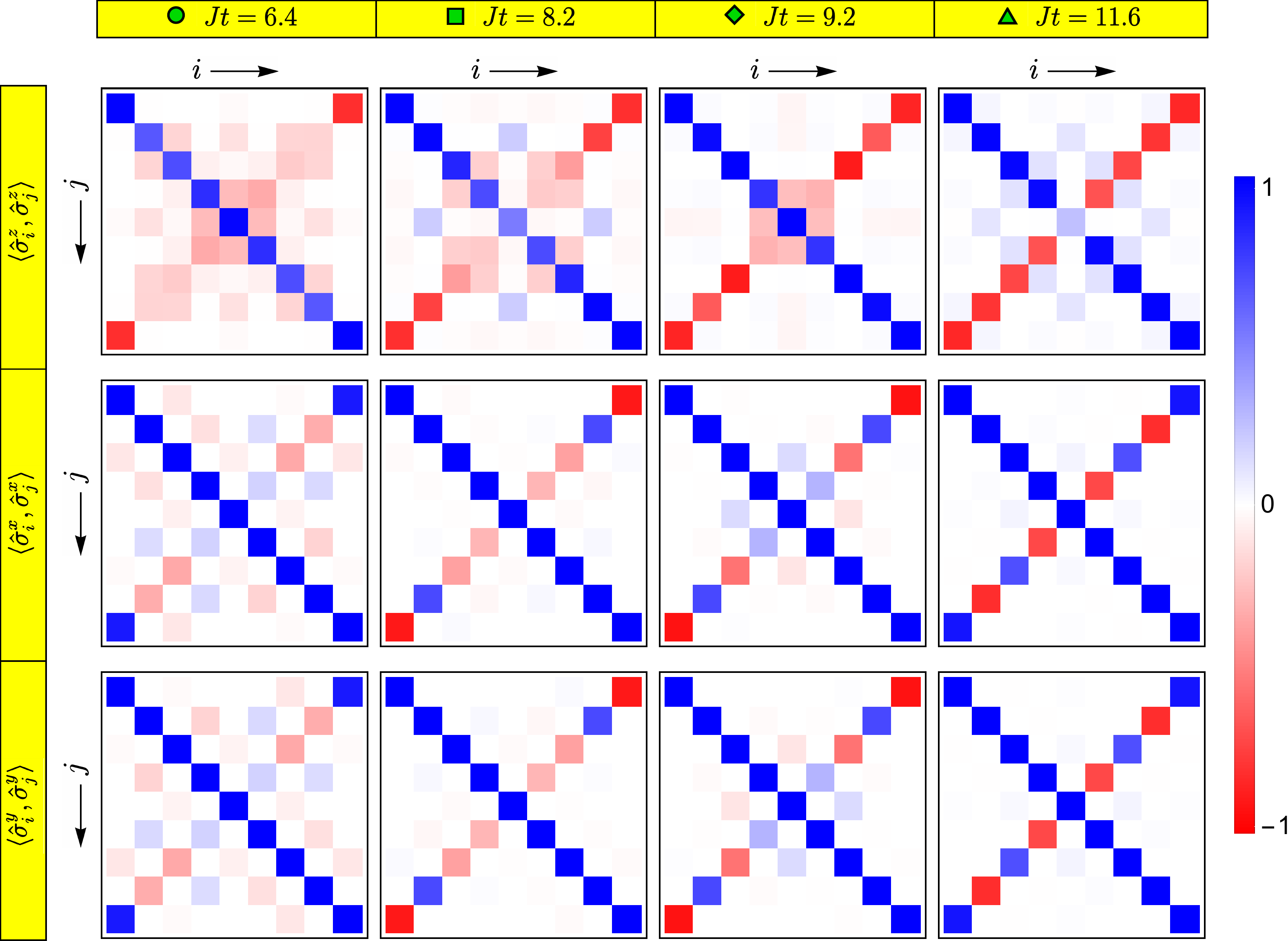}
	\centering
	\caption{{\bf Spin-spin correlations.} Time evolution of $\langle \hat{\sigma}^{\nu}_i , \hat{\sigma}^{\nu}_j \rangle := \langle \hat{\sigma}^{\nu}_i \hat{\sigma}^{\nu}_j \rangle - \langle \hat{\sigma}^{\nu}_i \rangle \langle \hat{\sigma}^{\nu}_j \rangle$ for a 9-site XX chain from exact diagonalization, reflecting the successive stacking of Bell pairs between mirror-conjugate sites in Figs.~\ref{fig:protocol_numerics}(c-f). Pulses are applied at times $Jt = 2.4, 5.0, 7.8, 10.2$, when the phase correlations $\langle \hat{\sigma}^x_i , \hat{\sigma}^x_{-i} \rangle$ and $\langle \hat{\sigma}^y_i , \hat{\sigma}^y_{-i} \rangle$ flip sign. Note that $\langle \hat{\sigma}^x_i \rangle = \langle \hat{\sigma}^y_i \rangle = 0$.
	}
	\label{suppfig:spincorr}
\end{figure}

\bigskip

\begin{figure}[H]
	\includegraphics[width=1\columnwidth]{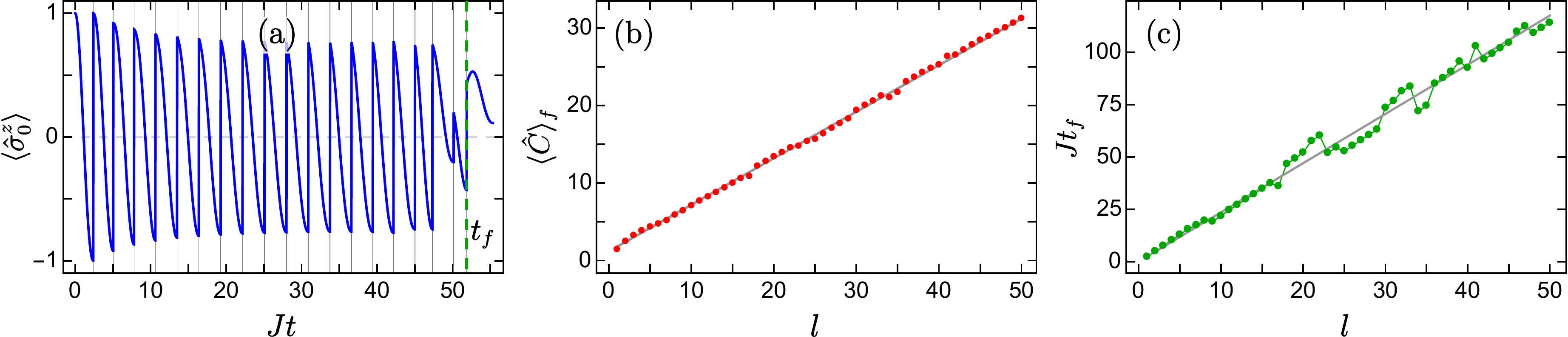}
	\centering
	\caption{{\bf Finite-size scaling.} (a) Evolution of the center-spin magnetization, $\langle \hat{\sigma}^z_0 \rangle$, during our protocol for 47 sites ($l=23$) from exact diagonalization. The last pulse is applied at time $t_f$, after which $\langle \hat{\sigma}^z_0 \rangle$ reaches a minimum above 0; continuing beyond this point does not yield significant gains. (b) Final value of $\langle \smash{\hat{C}} \rangle$ as a function of $l$, compared with $0.6(l+2)$ shown by the gray line. Note that $\langle \smash{\hat{C}} \rangle$ is bounded by $l+1/2$. (c) Duration as a function of $l$; gray line shows $2.35 l$. The sudden small fluctuations are due to variation in the number of pulses, which fluctuates around $0.8l$ for large $l$.
	}
	\label{suppfig:scaling}
\end{figure}

\clearpage

\begin{figure}
	\includegraphics[width=0.94\columnwidth]{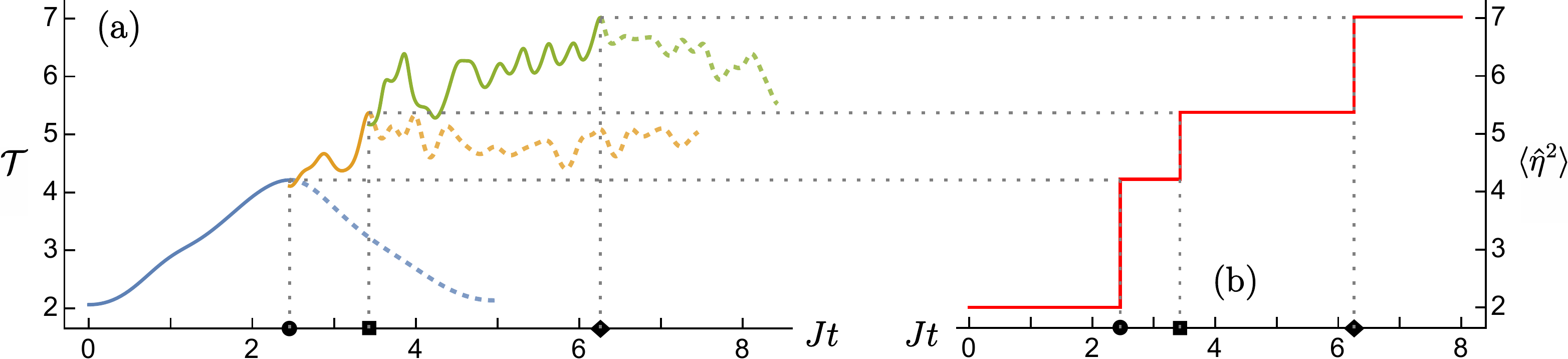}
	\centering
	\caption{{\bf Illustration of the protocol for generating $\eta$ pairs.} Evolution of a Fermi-Hubbard chain with 8 sites and $U/J=10$ using exact diagonalization. (a) Efficacy $\mathcal{T} := \smash{\langle \hat{\mathcal{J}} \hat{\eta}^2 \hat{\mathcal{J}} \rangle} / \smash{\langle \hat{\mathcal{J}}^2 \rangle}$ is monitored for a duration $U/(2J^2)$ and the next pulse of $\smash{\hat{\mathcal{J}}}$ is applied when $\mathcal{T}$ is maximum, shown by the dotted vertical lines. Solid curves show the resulting piecewise evolution. (b) $\langle \hat{\eta}^2 \rangle$ is conserved in between pulses and jumps to the instantaneous value of $\mathcal{T}$ after the application of each pulse.
	}
	\label{suppfig:etapairing}
\end{figure}

\endgroup

\end{document}